%% file: paren.tex
\input mn		%MNRAS ref MA6
				%astro-ph/0003479  pw njctf
\input mnextra \def\etal{et al}
\input psfig

\overfullrule=0pt
\newif\ifpsfiles\psfilestrue%false

\def\ex#1{\langle#1\rangle}

%\Referee
\begintopmatter
\title{The Age of the Solar Neighbourhood}

\author{James Binney$^1$, Walter Dehnen$^2$ and Gianpaolo Bertelli$^3$}
\affiliation{$^1$Theoretical Physics, University of Oxford, Oxford, OX1 3NP} 
\smallskip
\affiliation{$^2$Max-Planck-Institut für Astronomie, Königstuhl 17, D-69117 Heidelberg, Germany}
\smallskip
\affiliation{$^3$Università di Padova, Dipartimento di Astronomia, Vicolo dell'Osservatorio
5, 35122 Padova, Italy }
\shortauthor{J.J.\ Binney, W.~Dehnen and G.~Bertelli}
\shorttitle{Solar neighbourhood}

\abstract{High-quality Hipparcos data for a complete sample of nearly
$12\,000$ main-sequence and subgiant stars, together with Padua isochrones,
are used to constrain the star-formation history of the solar neigbourhood
and the processes that stochastically accelerate disk stars. The velocity
dispersion of a coeval group of stars is found to increase with time from
$\sim8\kms$ at birth as $t^{0.33}$. In the fits, the slope of the IMF near
$1\msun$ proves to be degenerate with the rate at which the star-formation
rate declines. If the slope of the IMF is to lie near Salpeter's value,
$-2.35$, the star-formation rate has to be very nearly constant. The age of
the solar neighbourhood is found to be $11.2\pm0.75\Gyr$ with remarkably
little sensitivity to variations in the assumed metallicity distribution of
old disk stars. This age is only a Gyr younger than the age of the oldest
globular clusters when the same isochrones and distance scale are employed.
It is compatible with current indications of the redshift of luminous galaxy
formation only if there is a large cosmological constant.  A younger age is
formally excluded because it provides a poor fit to the number density of red
stars. Since this density is subject to a significantly uncertain selection
function, ages as low as  $9\Gyr$ are plausible even though they lie outside
our formal error bars.}


\maketitle

\section{Introduction}

Observations at optical and near-infrared bands have now penetrated to the
redshifts $z\sim1$ at which it is widely believed that $L^*$ galaxies
formed. The extent and fragility of a thin galactic disk are such that a
disk like the one we inhabit would have been one of the last components to
form in a galaxy: the formation of any significant component interior to it
would at least have thickened it, and could easily have destroyed it
altogether. So it is interesting to ask `how old is the solar neighbourhood?'

The age of the solar neighbourhood has been estimated in several ways. One
method involves comparing measured isotope ratios with values predicted by
nucleosynthesis theory and radioactive decay rates. Such studies can only
place a lower limit on the age of the solar neighbourhood because the
samples studied are unlikely to be as old as the oldest objects in the solar
neighbourhood. This approach implies that the age of the solar neighbourhood
exceeds $9\Gyr$ (Morell, K\"allander \& Butcher, 1992) and could be as high
as $11\Gyr$.

A second approach to dating the solar neighbourhood involves identifying the
coolest, and therefore oldest, white dwarfs, and estimating their ages by
comparing their colours with those predicted by models of white-dwarf
cooling. This approach encounters two difficulties: (i) the theory of
white-dwarf cooling is complex and liable to error, and (ii) the oldest
white dwarfs are the least conspicuous and are easily missed. Several older
studies of cool white dwarfs yield ages in the range $6$ to $10\Gyr$
(Bergeron, Ruiz \& Legget, 1997), but recent models of cool white dwarfs
(Hansen, 1999; Saumon \& Jacobson, 1999)
suggest that the coolest objects may be bluer than previously thought and
may have been missed in older surveys.  It is possible that very old white
dwarfs have recently been identified through their proper motions in the
Hubble Deep Field (Ibata et al., 1999).

A third way to date the solar neighbourhood is to date individual F stars,
which have main-sequence lifetimes comparable to the age of the solar
neighbourhood. Since the age of a star of given colour and luminosity
depends sensitively on its metallicity, this technique requires accurate
metallicity estimates for the programme stars, which in practice causes the
available samples to be small and inadequately representative. Using the
sample of Edvardsson et al.\ (1993), Ng \& Bertelli (1998) conclude from
such a study that the solar neighbourhood contains stars that are at least 
$15\Gyr$ old.

A lower limit to the age of the disk can be set by determining the ages of
open clusters by fitting their colour--magnitude diagrams to theoretical
models. From six old open clusters and Padua isochrones Carraro, Girardi \&
Chiosi (1999) find the solar neighbourhood to be older than $10\Gyr$.

A fifth approach is to study the morphology of the subgiant branch in the
solar-neighbourhood colour--magnitude diagram. Since the solar
neighbourhood, unlike a globular cluster, contains stars that are widely
spread in both age and metallicity, the sequences in its colour-magnitude
diagram are broad and hard to characterize (see Fig.~1). None the less,
Jimenez, Flynn \& Kotoneva (1998) used this approach to place a lower limit
of $8\Gyr$ on the age of the solar neighbourhood. 

Here we date the solar neighbourhood by a new, sixth approach, in which the
local colour--magnitude diagram is combined with kinematic information.  It
has long been known that older populations of solar-neighbourhood stars have
larger velocity dispersions (Parenago, 1950; Roman, 1954), and this effect
gives rise to the phenomenon called Parenago's discontinuity: when stellar
velocity dispersion $\sigma$ is plotted against a quantity that is related
to mean stellar age, for example $B-V$, the slope of the graph changes
discontinuously near $B-V=0.6$ (Fig.~2), because redward of the
discontinuity one observes stars of every age, while blueward of it one
observes only the more recently formed stars. The Hipparcos Catalogue has
made it possible to quantify this phenomenon with unprecedented precision
(Binney \etal\ 1998; Dehnen \& Binney, 1998), and we show in this paper that
it enables us to obtain from the theory of stellar evolution an age for
the solar neighbourhood whose precision is comparable to those claimed for
the ages of globular clusters.

In addition to determining the solar neighbourhood's age, we  constrain
both the star-forming history of the solar neighbourhood, and the way in
which non-axisymmetric features in the Galactic potential cause
the random velocities of stars to increase with age.

In Section 2 we describe the data and the models. Section 3 presents the
results of the fits, Section 4 discusses their implications and Section 5
sums up.

\beginfigure1
\centerline{\psfig{file=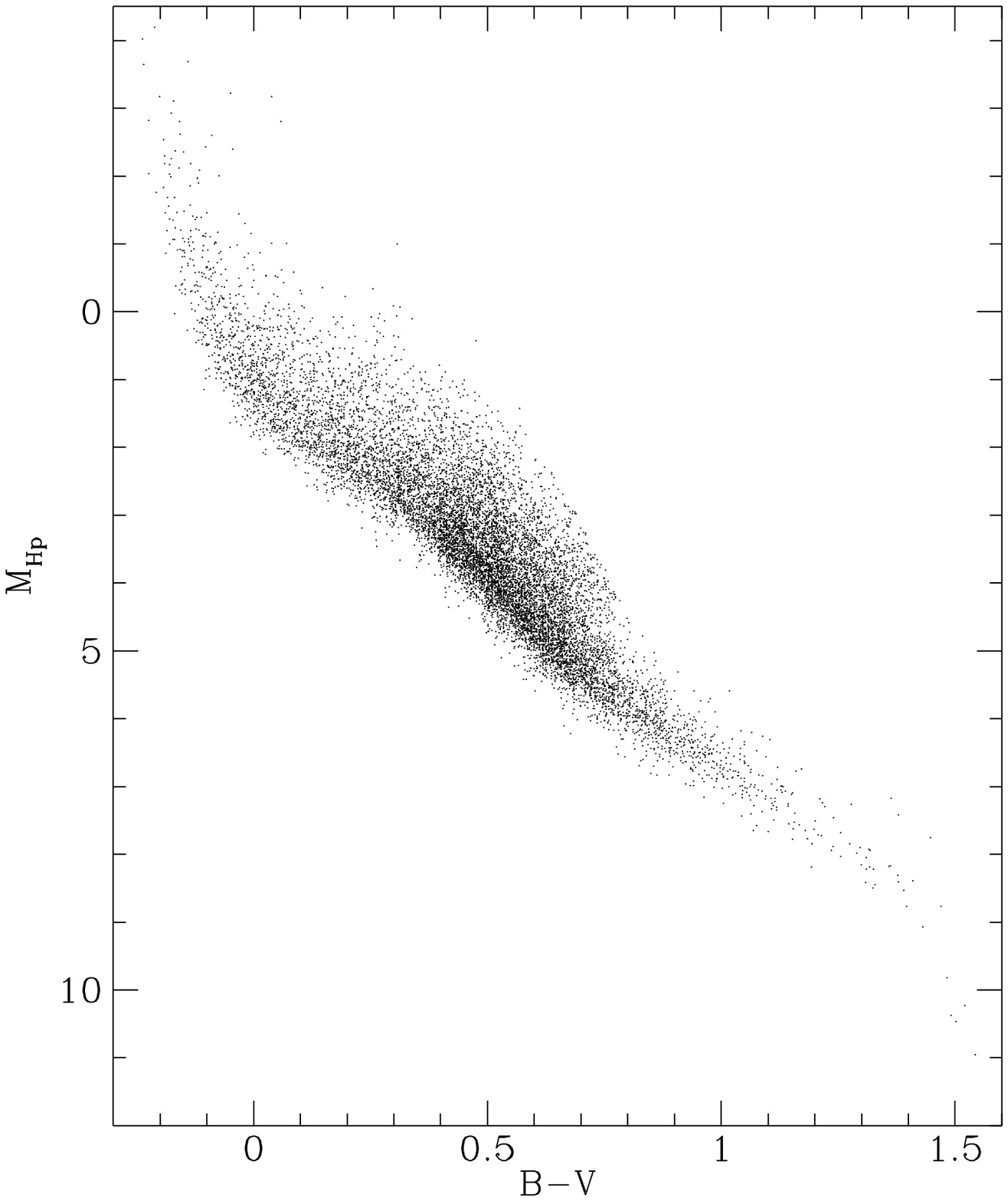,width=.8\hsize}}
\caption{{\bf Figure 1.} Colour--magnitude diagram of the sample.}
\endfigure

\beginfigure2
\centerline{\psfig{file=paren1a.ps,width=.8\hsize}}
\caption{{\bf Figure 2.} The dispersion in proper motions $S$ of
main-sequence stars as a function of dereddened $B-V$. The sharp change in gradient
near $B-V=0.6$ is Parenago's discontinuity.}
\endfigure

\beginfigure3
\centerline{\psfig{file=paren2.ps,width=\hsize}}
\caption{{\bf Figure 3.} The distributions, for $Z=0.014$, over age of stars at $B-V=0.55$ and
$B-V=0.7$. The solar-neighbourhood has been assumed to be $12\Gyr$ old and
the decay constant of the star-formation rate has been set to $t_0=10\Gyr$.}
\endfigure

\section{Data and models}

\subsection{The data}
Figs 1 and 2 show, for the photometrically complete sample of $11\,865$
Hipparcos stars that is described in Dehnen \& Binney (1998; hereafter Paper
I), the colour--magnitude diagram and a plot of velocity dispersion versus
$B-V$.  The quantity $S$ plotted in Fig.~2 is the rms dispersion in
proper-motion within each bin in $B-V$. Hence its relation to the principal
velocity dispersion $\sigma_R$, $\sigma_\phi$ and $\sigma_z$ depends on the
distribution of stars on the sky, which is not precisely uniform (see Fig.~2
of Dehnen 1998). However, for most bins we have
 $$
S\simeq\big[\frac23(\sigma_R^2+\sigma_\phi^2+\sigma_z^2)\big]^{1/2}.
\eqno\newe$$
 The data presented in Fig.~2 differ from those shown in Fig.~1 of Paper I
in two ways. First, each star has been dereddened by an amount proportional
to its distance:
\eqnam\BmVcorrect$$
E(B-V)=0.53\,(d/\!\kpc)
\eqno\newe$$
 from Binney \& Merrifield (1998), eq.~(3.66).  Second, the values of $S$
shown here are for a sliding window whose width varies such that there are
always 500 stars in the window.  Since a fresh point is plotted each time
100 stars have left the window, every fifth point is statistically
independent of it predecessors.

\subsection{The models}
We assume that star formation has at all times been characterized by a
power-law  IMF. Deviation of the actual IMF from power-law
form should be unimportant over the narrow range of stellar masses
($0.8\lta{\cal M}/M_\odot\lta2$) of interest here. 
The star-formation rate is assumed to decline 
exponentially in time with
characteristic rate constant $\gamma$. These assumptions imply that in a
volume-limited sample the
distribution of main-sequence stars over mass and age $\tau$ is
 $$
{\d^2N\over\d{\cal M}\d\tau}\propto
\cases{
{\cal M}^{-\alpha}\exp(\gamma\tau)&for $\tau<\tau_{\rm max}({\cal M})$,\cr
0&otherwise,}
\eqno\newe$$
 where $\tau_{\rm max}({\cal M})$ is the main-sequence lifetime of a star of
initial mass ${\cal M}$. 

Our sample is volume-limited for stars more luminous than $\sim5\lsun$
because these stars lie above the sample's approximate magnitude limit
($V\sim8$) at the distance, $d\sim100\pc$ at which their parallaxes become
too unreliable to be included in the sample. For less luminous stars, the
sample is effectively apparent-magnitude limited and equation \refe1) has to
be modified by introducing a factor $\phi$ that is proportional to the volume
surveyed.  Clearly, in the magnitude-limited regime, $\phi$ is proportional to
$L^{3/2}$. Hence we model the distribution of stars in the sample to be
 $$
{\d^2N\over\d{\cal M}\d\tau}\propto
\cases{
\phi{\cal M}^{-\alpha}\exp(\gamma\tau)&for $\tau<\tau_{\rm max}({\cal M})$,\cr
0&otherwise,}
\eqno\firste$$
 where 
$$
\phi({\cal M},t)=\cases{(L/5\lsun)^{3/2}&for $L<5\lsun$\cr1&otherwise.}
\eqno\laste b$$

We are interested in the distribution over age of
stars of a given colour $C$. This is given by
$$\eqalign{
{\d N\over\d\tau}\bigg|_{C_0}&\propto
\int\d{\cal M}\,\delta[C({\cal M},\tau)-C_0]{\d^2N\over\d{\cal M}\d\tau}\cr
&\propto {\phi{\cal M}^{-\alpha}\exp(\gamma\tau)\over |(\p C/\p{\cal M})_\tau|}
}\eqno\newe$$
 Thus, to proceed further we require $(\p C/\p{\cal M})_\tau$, the rate of
change of colour with mass at a given age, which is in principle readily obtained from
the appropriate isochrone. We have used the Padua isochrones of Bertelli
\etal\ (1994) for metallicities $Z=0.02$, $Z=0.014$ and $0.008$. 
Since the data are restricted to stars on or near the ZAMS,
we cut each isochrone off above the point at which it is $1.8\,$mag more
luminous than the ZAMS at the same colour.

Several technical problems arise in the determination of  $(\p C/\p{\cal
M})_\tau$ from a set of isochrones. First, isochrones are available only at
certain times. From these we  estimate the isochrone for any given time by
interpolation between the isochrones for the closest time before and the
closest time after the desired time. Let $C^-({\cal M^-})$ be the colour for
mass ${\cal M}^-$ at the earlier time. Then we seek the colour $C^+$ 
at the later time of the star that has mass
$$
{\cal M}^+={\cal M}_{\rm min}+
{{\cal M}^--{\cal M}^-_{\rm min}\over{\cal M}^-_{\rm max}-{\cal M}_{\rm min}}
({\cal M}^+_{\rm max}-{\cal M}_{\rm min}),
\eqno\newe$$
where ${\cal M_{\rm min}}$ is the  mass at which all isochrones start and 
${\cal M^\pm_{\rm max}}$ is the mass at which the earlier/later isochrone is
truncated. Having located $C^\pm$ and ${\cal M^\pm}$ we linearly interpolate
between them in time to obtain $C({\cal M})$ at the desired time.

Once an isochrone has been constructed for a given time, we normally obtain
the average of $|\p C/\p{\cal M}|^{-1}$ over each colour bin by
analytically integrating  over the quadratic in ${\cal M}$ that passes
through three appropriate points on the isochrone; this procedure yields a
finite value even when $(\p C/\p{\cal M})$ vanishes in the range of
integration. Unfortunately, isochrones often show poorly-sampled sharp
turnings. At such points the quadratic that passes through the data points
can be unphysical and we have deemed it wiser to obtain the average of $|\p
C/\p{\cal M}|^{-1}$ from linear approximations to $C({\cal M})$.

Fig.~3 shows, for metallicity $Z=0.014$ ($\hbox{[Fe/H]}=-0.155$), the
distributions over age for two colours, namely $B-V=0.55$ and $B-V=0.7$. At
the bluer colour, all stars are younger than $6.1\Gyr$, while at the redder
colour, which lies redward of Parenago's discontinuity, there are stars of
every age up to the assumed age of the disk, $11\Gyr$. Both distributions
are sharply peaked in age.  The step at age $\tau=4.2\Gyr$ in the
distribution for $B-V=0.7$ arises because isochrones older that $4.2\Gyr$
cross $B-V=0.7$ twice, once on the main sequence and once on the subgiant
branch. The peaks at $\tau=4.5\Gyr$ in the distribution for $B-V=0.55$ are
associated with times at which the main-sequence turnoff lies near
$B-V=0.55$ and $(\p C/\p{\cal M})$ vanishes at this colour.

The age distribution at a given colour depends strongly on the metallicity
of the isochrones employed. In reality the metallicities of
solar-neighbourhood stars cover a band of finite width. The observational
characterization of this band has been hampered by the want of an unbiased
sample of disk stars with accurately determined metallicities. However, data
from Pagel \& Patchett (1975), Rana \& Basu (1990), Rocha-Pinto \& Maciel
(1996) and Flynn \& Morell (1998) suggest that the band is centred on
$\hbox{[Fe/H]}\simeq-0.15$ and is $\sim0.5\,$dex wide (cf.\ Fig.~7 of Flynn
\& Morell). We crudely simulate the band with a sum of three Dirac
delta-functions in metallicity at $\hbox{[Fe/H]}=-0.4,\,-0.15$ and $0$ with
respective weights $0.25,\,0.5$ and $0.25$. Table 1 gives, for this
metallicity mixture and a constant star-formation rate,
the distribution in age of the stars that would be
picked up in each of three colour bands.  The bands, which are all
$0.1\,$mag wide in $B-V$, lie, respectively, blueward of Parenago's
discontinuity (central $B-V=0.4$), across the discontinuity (central
$B-V=0.6$) and redward of it (central $B-V=0.8$).

\begintable1
\caption{{\bf Table 1.} The distribution over age $\tau$ of 
stars in three colour bands.
Each band is $0.1\,$mag wide in $B-V$.}
\halign to .8\hsize{
\hfil$#$&\hfil$#$&\hfil$#$&\hfil$#$\cr
\noalign{\vskip4pt\hrule\vskip4pt}
\tau&\quad(0.35-0.45)&\quad(0.55-0.65)&\quad(0.75-0.85)\cr
\noalign{\vskip4pt\hrule\vskip4pt}
    0.50&  23.30&   4.79&   1.36\cr
    1.00&  30.40&   4.95&   1.35\cr
    1.50&  71.36&   5.17&   1.35\cr
    2.00&  67.05&   5.83&   1.35\cr
    2.50&  18.98&   9.70&   1.35\cr
    3.00&  10.66&  13.67&   1.35\cr
    3.50&   8.21&  15.61&   1.35\cr
    4.00&   1.03&  13.84&   1.31\cr
    4.50&   0.03&  17.05&   1.34\cr
    5.00&   0.00&  24.06&   1.37\cr
    5.50&   0.00&  21.55&   1.50\cr
    6.00&   0.00&  21.25&   1.59\cr
    6.50&   0.00&  18.44&   1.61\cr
    7.00&   0.00&  16.68&   1.60\cr
    7.50&   0.00&  11.57&   1.61\cr
    8.00&   0.00&  24.20&   1.62\cr
    8.50&   0.00&  16.98&   1.88\cr
    9.00&   0.00&  13.97&   2.06\cr
    9.50&   0.00&   9.91&   2.18\cr
   10.00&   0.00&   8.82&   2.33\cr
\noalign{\vskip4pt\hrule}
}
\endtable

\beginfigure*4
\centerline{\psfig{file=paren3b.ps,width=.8\hsize}}
\caption{{\bf Figure 4.} Top panel: data from Fig.~2 together with two model
fits. The full (dashed) curve is for the model whose parameters are given in the
top (bottom) row of Table 2.  The solar-neighbourhood metallicity distribution is
taken to be the sum of three Dirac delta functions at $Z=0.008,\,Z=0.014$
and $Z=0.02$ with weights $0.25,\,0.5$ and $0.25$, respectively. Lower
panel: the corresponding fits to the number of stars in each colour bin.  }
\endfigure

The theory of stellar scattering has been examined many times; we refer the
reader to \S7.5 of Binney \& Tremaine (1987) for an elementary discussion
and to Jenkins (1992) an up-top-date treatment and references to the
literature. In the simplest models, the stellar velocity dispersion
increases as some power of time, $\sigma\sim t^\beta$, with
$0.2\lta\beta<0.5$, and we assume such a power-law dependence here.
Specifically, 
we take the velocity dispersion $S$ of a group of stars of known age
distribution to be given by
 \eqnam\accelform$$
S^2={\int_0^{\tau_{\rm max}}\d\tau\,(\d N/\d\tau)
\{v_{10}[(\tau+\tau_1)/(10\Gyr+\tau_1)]^\beta\}^2
\over
\int_0^{\tau_{\rm max}}\d\tau\,(\d N/\d\tau)}
\eqno\newe$$
 In this formula $\tau_{\rm max}$ is the age of the solar neighbourhood,
$\tau_1$ controls the random velocity of stars at birth, and $v_{10}$ and
$\beta$ characterize the efficiency of stellar acceleration.

\section{Fits to the data}
The data are of two types. The data plotted in Fig.~2 constrain the
parameters, $\beta,\,v_{10},\,\tau_1$ and $\tau_{\rm max}$ that characterize
the stochastic acceleration process, but give little leverage on the
parameters $\alpha$ and $\gamma$ that parameterize the disk's star-formation
history. The last two parameters are, however, strongly constrained by the
number of stars in each colour bin. Hence, we use the Levenberg--Marquardt
non-linear least-squares algorithm (Press \etal\ 1986) to minimize the sum
of the $\chi^2$ values of the model's fits to both the numbers of stars and
the velocity dispersion of each colour bin.  

Only every fifth of the data points plotted in Fig.~2 is statistically
independent, so we have fitted to only a subset of the data. When only every
fifth data point was employed, to ensure that all the data points used were
statistically independent and $\chi^2$-values were easily interpreted, the
results proved to depend significantly on {\it which\/} subset of the data
were used. More nearly consistent results are obtained when every third data
point is employed, so that every data point is statistically independent of
its second-nearest neighbours.  The results reported here are for this case.
Hence, $\chi^2$-values can be expected to be lower than those one would
obtain for statistically independent data points.

\begintable*2
\caption{{\bf Table 2.} Fitted values of the parameters for four models. The
first three rows show fits for different
weightings of the metallicity components $\hbox{[Fe/H]}=-0.4,\,-0.15,\,0$: 
from top to bottom the weightings are (0.25,0.5,0.25), (0.333,0.333,0.333)
and (0.2,0.6,0.2).
The fit given by the first row is plotted in Fig.~4. The last row shows the
result of fixing $\tau_{\rm max}=9\Gyr$ with metallicity weightings
 (0.25,0.5,0.25).}
\line{\hfil\vbox{\halign to 11.7cm{%
\hfil$#$\hfil&\quad \hfil$#$\hfil&\quad \hfil$#$\hfil&\quad \hfil$#$\hfil&\quad \hfil$#$\hfil&\quad
\hfil$#$\hfil\cr
\noalign{\vskip4pt\hrule\vskip4pt}
\alpha&\beta&\gamma/\hbox{Gyr}^{-1}&\tau_{\rm max}/\hbox{Gyr}&\tau_1/\hbox{Gyr}&v_{10}/\!\kms\cr	
\noalign{\vskip4pt\hrule\vskip4pt}
 -2.26\pm0.48& 0.328\pm0.021& -0.006\pm0.046& 11.16\pm0.77& 0.030\pm0.072& 58.1\pm1.4\cr
 -2.58\pm0.47& 0.327\pm0.022& -0.035\pm0.044& 11.80\pm0.39& 0.039\pm0.074& 58.0\pm1.5\cr
 -2.14\pm0.48& 0.322\pm0.093& -0.000\pm0.045& 11.19\pm0.74& 0.006\pm0.065& 57.8\pm1.3\cr
\noalign{\vskip4pt}
 -1.74\pm0.47& 0.357\pm0.024&  0.075\pm0.046& 9.000       & 0.107\pm0.093& 59.9\pm1.5\cr
\noalign{\vskip4pt\hrule\vskip4pt}
}}\hfil}
\endtable

\begintable3
\caption{{\bf Table 3.} The normalized covariance matrix, $\ex{\delta
p_i\delta p_j}/(\sigma_i\sigma_j)$, of the model parameters for
the six-parameter fit shown by the full curves in Fig.~4 and the first row of Table~2.}
\line{\hfil\vbox{\halign to 6.5cm{%
$#$\hfil\quad&\hfil$#$\quad &\hfil$#$\quad &\hfil$#$\quad &\hfil$#$\quad &\hfil$#$\cr
\noalign{\vskip4pt\hrule\vskip4pt}
&\alpha\hfil&\beta\hfil&\gamma\hfil&\tau_{\rm max}\hfil&\tau_1\hfil\cr
\noalign{\vskip4pt\hrule\vskip4pt}
\beta		  &-0.338\cr
\gamma		  & 0.908& -0.322\cr
\tau_{\rm max}&-0.340& -0.240& -0.491 \cr
\tau_{\rm 1}  &-0.200&  0.910& -0.173& -0.203 \cr
v_{10}		  &-0.570&  0.850& -0.581& -0.156&  0.615 \cr
\noalign{\vskip4pt\hrule}
}}\hfil}
\endtable

%6-p fit z=0.2,0.5,0.2; chisq=0.78
% -2.260\pm0.48  0.328\pm0.021 -0.006\pm0.046 11.161\pm0.77  0.030\pm0.072 58.124\pm1.4
% 0.482E+00
%-0.338E+00  0.210E-01
% 0.908E+00 -0.322E+00  0.460E-01
%-0.340E+00 -0.240E+00 -0.491E+00  0.771E+00
%-0.200E+00  0.910E+00 -0.173E+00 -0.203E+00  0.720E-01
%-0.570E+00  0.850E+00 -0.581E+00 -0.156E+00  0.615E+00  0.140E+01

%6-p fit z=0.33,0.33,0.33; chisq=0.70
% -2.579\pm0.47  0.327\pm0.022 -0.035\pm0.044 11.803\pm0.39  0.039\pm0.074 58.048\pm1.5
% 0.469E+00
%-0.457E+00  0.223E-01
% 0.905E+00 -0.472E+00  0.438E-01
%-0.246E+00 -0.175E+00 -0.376E+00  0.391E+00
%-0.298E+00  0.912E+00 -0.291E+00 -0.156E+00  0.742E-01
%-0.661E+00  0.864E+00 -0.709E+00 -0.103E+00  0.635E+00  0.155E+01

%6-p fit z=0.2,0.6,0.2; chisq=0.88
% -2.139\pm0.476E  0.322\pm0.093  0.000\pm0.045 11.192\pm0.74  0.006\pm0.065 57.767\pm1.3
% 0.476E+00
%-0.326E+00  0.193E-01
% 0.907E+00 -0.314E+00  0.450E-01
%-0.338E+00 -0.233E+00 -0.490E+00  0.742E+00
%-0.188E+00  0.900E+00 -0.166E+00 -0.189E+00  0.647E-01
%-0.556E+00  0.846E+00 -0.566E+00 -0.157E+00  0.597E+00  0.133E+01

The first three rows of Table~2 show the parameters of three fits to the
data that differ in the assumed metallicity composition of the solar
neighbourhood: the first row is for weightings $(0.25,0.5,0.25)$ of the
three metallicity components, while the second and third rows are for
$(0.333,0.333,0.333)$ and $(0.2,0.6,0.2)$, respectively. The differences
between values of a given parameter for different assumed metallicities are
comparable to or smaller than the formal error in that parameter. Thus, our
results are effectively independent of reasonable assumptions about the
metallicity of the solar neighbourhood.

The formal error in $\alpha$ is quite large ($\sim25\%$). Moreover, $\alpha$
is strongly correlated with the rate parameter $\gamma$: Table 3 shows that
the correlation coefficient between these two parameters is nearly unity.
This is because increasing $\gamma$ diminishes the current star-formation
rate relative to its historical average, and this makes blue stars
relatively scarce. This effect can be effectively cancelled by increasing
$\alpha$ and thus flattening the IMF. The values of $\alpha$ given in the
first three rows of Table~2 are consistent with Salpeter's classical value,
$\alpha=-2.35$.

Tables 2 and 3 show that the formal error in $\beta$ is small ($\sim0.02$),
but $\beta$ is highly correlated with $\tau_1$. The formal error in the age
of the solar neighbourhood, $\tau_{\rm max}$ is $\sim0.75\Gyr$ and this
parameter is not particularly strongly correlated with any other. 

The age, $11.2\pm0.75\Gyr$, to which the first three rows of Table~2 point
is older than recent models of globular clusters and cosmological
developments would lead one to expect (see below). So it is interesting to
fix $\tau_{\rm max}$ at a younger age and see how well the data can be
fitted by varying the other parameters. The bottom row of Table~2 and the
dashed curves in Fig.~4 show the result of this exercise when we set
$\tau_{\rm max}=9\Gyr$. The value of $\chi^2$ per degree of freedom has
risen from $0.78$ when $\tau_{\rm max}$ is varied to $0.95$. Fig.~4 shows
that this increase in $\chi^2$ is attributable to a deterioration in the fit
to the observed density of red stars. In fact, the model with the smaller
age gives a slightly better fit to the velocity data plotted in the upper
panel of Fig.~4, especially in the crucial region around the discontinuity.
Since the fits to the stellar number densities are sensitive to our rather
uncertain selection function, the model with $\tau_{\rm max}=9\Gyr$ should
be considered at least as plausible as the models with older ages.

The deficit of red stars in the model with $\tau_{\rm max}=9\Gyr$ arises
from two factors: the model has a flat IMF, $\alpha=-1.74$, and
a star-formation rate that increases with time.

\section{Discussion}

Our approach to the determination of the age of the solar neighbourhood has
much in common with the work of Jimenez, Flynn \& Kotoneva (1998): we both use
samples of Hipparcos stars and stellar evolution as the clock. However,
since the sub-giant branch is not evident in Fig.~1, our technique is
radically different from that of Jimenez et al., which focused on the
morphology of the sub-giant branch. Our results show that by including
kinematic information, which provides a means of distinguishing old from
young stars, at least in a statistical sense, we can overcome the difficulty
that one encounters when one attempts to transfer from globular clusters to
the solar neighbourhood a dating technique that relies on the sharpness of
sequences in the colour--magnitude diagrams of globular clusters.
Our final age estimate, $11.2\pm0.75\Gyr$, is compatible with the constraint
that Jimenez et al.\ derive, $\tau_{\rm max}>8\Gyr$. 

In addition to its cited $0.75\Gyr$ formal error, our estimate of the age of
the solar neighbourhood is subject to any systematic error in the isochrones
we have employed. Great strides have been taken by stellar evolution models
within the last decade, largely as a result of upgrades to the atomic
physics used. In particular, there is now good agreement between observation
and theory in the pulsation periods of variable stars (Chiosi et al.\ 1992).
Also, models produced by different groups are in excellent agreement with
one another, especially when convective overshoot is unimportant, as is in
the case of main-sequence stars older than $\sim6\Gyr$. The remaining
discrepancies between the isochrones of different groups are principally due
to different transformations between effective temperature and colours such
as $B-V$. In this connection it is noteworthy that displacing our
solar-metallicity isochrones for $8,\,11$ and $15\Gyr$ by $0.05\,$mag to the
blue yields almost perfect agreement with the isochrones of Jimenez et al.\
near the crucial turn-off. When the velocity-dispersion data are fitted with
such displaced isochrones, $\tau_{\rm max}$ increases by 3 to $4\Gyr$
depending on the assumed metallicity distribution.

The difference between our solar-neighbourhood age and the ages of globular
clusters should be less subject to systematic error than an absolute age.
Gratton et al.\ (1997) have determined the ages of globular clusters by
fitting the absolute magnitude, $M_V({\rm TO})$ of the main-sequence turnoff
to the isochrones of different groups. The Padua
isochrones yielded the youngest ages, ranging up to $12.4\Gyr$ for NGC~6341.
Hence, by this reckoning, the solar neighbourhood is only one Gyr younger
than an old globular cluster. Weight is lent to this conclusion by the fact
that Gratton et al.\ determined $M_V({\rm TO})$ by forcing the cluster
main-sequences to pass through the locations of sub-dwarfs of appropriate
metallicity for which Hipparcos had determined a parallax. Hence, we are
comparing ages that are based on the same isochrones and the same distance
scale. If the comparison is misleading, it will be because our age of the
solar neighbourhood relies on the colour calibration of the Padua
isochrones, whereas that of Gratton et al.\ was designed to be largely free
from this calibration. Note, however, that shifting the calibration in the
sense suggested by the work of Jimenez would make the solar neighbourhood
older than the oldest globular clusters.

As Table~2 illustrates, the parameters $\beta$, $\tau_{\rm max}$, $\tau_1$
and $v_{10}$ that characterize the age of the disk and the stochastic
acceleration of stars, are remarkably well determined by the models. The
exponent $\beta$ lies in the range expected if the acceleration process is
primarily driven by spiral structure and modulated by stars scattering off
giant molecular clouds (Jenkins 1992). Our allowed range,
$\beta=0.33\pm0.03$, excludes the value $\beta=0.5$ that has been advocated
by Wielen (Wielen, 1977; Wielen, Fuchs \& Dettbarn 1996).  Notice that for
$\beta=\frac13$ and $v_{10}=58\kms$, equation \accelform) predicts that the
velocity dispersion of stars at birth is $8.4\kms$ for $\tau_1=0.03\Gyr$.
For comparison, spiral structure endows molecular clouds with non-circular
velocities $\sim7\kms$ (Malhotra, 1994), while the internal velocity
dispersions within molecular clouds are $\sim3\kms$ (Solomon \& Rivolo,
1989). Hence $8.4\kms$ implies that during formation stars acquire slightly
larger random velocities than the gas from which they formed had.

The degeneracy between $\alpha$ and $\gamma$ that is quantified by Table 3
has been thoroughly discussed by Haywood, Robin \& Cr\'ez\'e (1997a). These
authors point out that $\alpha$ and $\gamma$ are connected not only with one
another, but also with the vertical structure of the disk, because older and
less luminous stars form a thicker disk than young and luminous stars, so
the latter tend to be over-represented in solar-neighbourhood samples.  In
fact, the value of $\alpha$ returned by our fits is sensitive to the
selection function $\phi$ that we adopt, and this is not as well defined as
we would like because, the sample, in common with virtually all Hipparcos
samples, involves complex selection criteria. In particular, our primary
selection criterion is by relative parallax error.  This error is primarily
a function of distance, but it has non-negligible dependence on both
ecliptic latitude and apparent magnitude.  Early-type stars will tend to be
included to greater distances than late-type stars.

\beginfigure5
\centerline{\psfig{file=cosmos.ps,width=.8\hsize}}
\caption{{\bf Figure 5.} The full curve shows the relationship between redshift and time for
$H_0=65\kms\Mpc^{-1}$, $\Omega_{\rm matter}=0.3$ and $\Omega_\Lambda=0.7$.
The dashed curve is for the same parameter values except $\Omega_\Lambda=0$.
The $1\sigma$-error band of our determination of the
age of the solar neighbourhood is shaded.}
\endfigure

In a second paper Haywood, Robin \& Cr\'ez\'e (1997b) exploit the connection
with the disk's vertical structure by constraining their models of the
disk's evolution with both local data and faint star counts at the Galactic
poles. They find that these fits favour a flat IMF and a constant or
increasing star-formation rate, very much in line with the result of our
fits to the data. This conclusion is in marked contrast with
that reached by Scalo (1986), who found that in the relevant mass range
($0.8\msun\lta{\cal M}\lta2\msun$) the slope of the IMF is flattening from
$\alpha\sim-3.3$ at ${\cal M}>\msun$ to $\alpha\sim-1.8$ at ${\cal
M}<\msun$, and with the conclusion of Kroupa, Tout \& Gilmore (1993),
who found  that the IMF is as steep as $\alpha=-4.5$ for ${\cal M}>\msun$ and
has $\alpha\sim-2.2$ below $\!\msun$. Unfortunately, the strength of the
correlation between $\alpha$ and $\gamma$ in our fits prevents  us from
throwing significant weight behind one side or the other in this
controversy. 

We should also remark that the physical meaning of the parameters $\alpha$
and $\gamma$ used here may differ slightly from the meaning of the
corresponding parameters in other studies. Since stars can diffuse into and
out of the volume that was surveyed by Hipparcos, the derivative with
respect to age of the density of stars of a given type will strictly differ
from the rate at which such stars formed at the corresponding time in the
past. For example, the thickness of the disk increases with age, so the
density of objects near the plane, where Hipparcos observed, will increase
with $\tau$ less rapidly than the true star-formation rate did. Radial
diffusion of stars could have a similar effect, and these effects
will slightly reduce the magnitude of the measured value of $\gamma$.

There is currently much interest in determining the redshift at which
galactic disks such as our own formed. What redshift does an age of
$11.2\Gyr$ correspond to? The full curve in Fig.~5 shows the redshift--age
relationship in the currently popular $\Omega_{\rm matter}=0.3$,
$\Omega_\Lambda=0.7$ cosmology for $H_0=65\kms\Mpc$, together with the
formal $1\sigma$-error band of our age determination. The lower edge of this
band crosses the age--$z$ relation at $z=1.6$, which is on the high side of
current estimates of the redshift at which $L^*$ disks such as that of the
Milky Way formed (Baugh et al., 1998; Madau, Pozzetti \& Dickinson, 1998).
The dashed curve shows the redshift--age relationship for the same Hubble
constant and $\Omega_{\rm matter}$ but vanishing cosmological constant. This
curve enters our possible age band at $z=3.1$, which considerably exceeds
the redshift at which $L^*$ disks are believed to form. Hence, our age
estimate is compatible with $H_0=65\kms\Mpc^{-1}$ only if the cosmological
constant is currently important. If $H_0$ were as small as
$55\kms\Mpc^{-1}$, the redshift--age relation would be compatible with our
age estimate for both $\Omega_\Lambda=0.7$ (when $z_{\rm formation}\gta1.1$)
and $\Omega_\Lambda=0$ (when $z_{\rm formation}\gta1.7$). 

Fig.~5 shows that combining the evidence that $L^*$ disks formed at $z\sim1$
with either of  the cosmological models that have $H_0=65\kms\Mpc^{-1}$
leads to the conclusion that the solar neighbourhood should be $\sim9\Gyr$
old. Fig.~4 shows that such a low value of $\tau_{\rm max}$ is formally
excluded because it yields a poor fit to the number density of stars, even
though it provides an excellent fit to the velocity data. Since the model
stellar densities are subject to a significantly uncertain selection
function, a low value of $\tau_{\rm max}$ that is compatible with
$H_0=65\kms\Mpc^{-1}$ and $\Omega_\Lambda=0$ should not be regarded as
excluded despite the formal error bars on our age determination.

\section{conclusions}

By fitting a set of high-quality Hipparcos data with state-of-the-art
isochrones we have jointly constrained the star-formation history of the
solar neighbourhood and the mechanism that stochastically accelerates stars.
The value we obtain for the slope of the IMF near $1\msun$ is compatible
with the classical Salpeter value, but the uncertainty of this parameter is
large because an increase or decrease of it can be almost perfectly
compensated by a change in the time constant that quantifies the decline in
the star formation rate. If the slope of the IMF is to lie near Salpeter's
value, the star formation rate has to be almost constant in time. A good fit
to the data is obtained by assuming that the velocity dispersion of a coeval
group of stars increases from $\sim8\kms$ as $t^\beta$ with
$\beta=0.33\pm0.02$. The age of the solar neighbourhood is found to be
$11.2\pm0.75\Gyr$, which is slightly older than currently popular values of the
cosmological constants and likely redshifts for the formation of massive
stellar disks would predict. A ounger age for the solar neighbourhood,
$\sim9\Gyr$ that is suggested with cosmological considerations is formally
excluded because it provides a poor fit the the number density of red stars.
Since this number density is subject to a significantly uncertain selection
function, such young ages cannot be considered to have been securely excluded.

\section*{Acknowledgments}

We thank M.G. Edmunds for valuable comments on the first version of this
paper.

\bye

%% file: mnextra.tex
\hoffset=-.5cm
\voffset=.5cm
% Bold math italic
\font\fivebmi=cmmib6
\font\sixbmi=cmmib6	\skewchar\sixbmi='177
\font\ninebmi=cmmib10 at 9pt 	\skewchar\ninebmi='177
\newfam\bmifam
\textfont\bmifam=\ninebmi
\scriptfont\bmifam=\sixbmi
\scriptscriptfont\bmifam=\fivebmi

%-------------redefine greek chars to be used as bold etc.
\mathchardef\alpha="710B
\mathchardef\beta="710C
\mathchardef\gamma="710D
\mathchardef\delta="710E
\mathchardef\epsilon="710F
\mathchardef\zeta="7110
\mathchardef\eta="7111
\mathchardef\theta="7112
\mathchardef\iota="7113
\mathchardef\kappa="7114
\mathchardef\lambda="7115
\mathchardef\mu="7116
\mathchardef\nu="7117
\mathchardef\xi="7118
\mathchardef\pi="7119
\mathchardef\rho="711A
\mathchardef\sigma="711B
\mathchardef\tau="711C
\mathchardef\upsilon="711D
\mathchardef\phi="711E
\mathchardef\chi="711F
\mathchardef\psi="7120
\mathchardef\omega="7121
\mathchardef\varepsilon="7122
\mathchardef\vartheta="7123
\mathchardef\varpi="7124
\mathchardef\varrho="7125
\mathchardef\varsigma="7126
\mathchardef\varphi="7127

\def\chaphead{}
\newcount\eqnumber
\eqnumber=1

\def\today{\ifcase\month\or
 January\or February\or March\or April\or May\or June\or
 July\or August\or September\or October\or November\or December\fi
 \space\number\day, \number\year}

%to name an equation, place command "\eqnam{\Poisson}" before equation, and
%thereafter "equation \Poisson)" will generate the proper equation number.
\def\eqnam#1{\xdef#1{(\chaphead\the\eqnumber}}

\def\newe{(\hbox{\chaphead\the\eqnumber})\global\advance\eqnumber by 1}
\def\firste{(\hbox{\chaphead\the\eqnumber a})\global\advance\eqnumber by 1}
\def\laste#1{\advance\eqnumber by -1%
	(\hbox{\chaphead\the\eqnumber #1})\advance\eqnumber by 1}

%to refer to an equation which is 5 equations back, write "equation \refe5)"
\def\refe#1{\advance\eqnumber by -#1 (\chaphead\the\eqnumber
     \advance\eqnumber by #1 }

			%fourth=4\th\ etc

\def\i{\relax\ifmmode{\rm i}\else\char16\fi}

             %for angular measure in degrees
\def\frac#1#2{{\textstyle{#1\over#2}}}
\def\p{\partial}
\def\d{{\rm d}}
\def\dddot#1{\ddot#1\kern-1.4pt\dot{\phantom{#1}}\kern-3pt}

 %error function
 %hyperbolic sec
 %hyperbolic csc
 %arc hyperbolic sin
 %arc hyperbolic cos
 %arc hyperbolic tan
 %arc hyperbolic cot
 %arc hyperbolic sec
 %arc hyperbolic csc

 %arc cot
 %arc sec
 %arc csc

          %spherical harmonic
   %spherical harmonic primed
                            %script E
                   %real part
                   %imaginary part
\def\spose#1{\hbox to 0pt{#1\hss}}

\def\=#1{\overline{#1}}

%\lta and \gta produce > and < signs with twiddle underneath
\def\lta{\mathrel{\spose{\lower 3pt\hbox{$\mathchar"218$}}
     \raise 2.0pt\hbox{$\mathchar"13C$}}}
\def\gta{\mathrel{\spose{\lower 3pt\hbox{$\mathchar"218$}}
     \raise 2.0pt\hbox{$\mathchar"13E$}}}

\def\kms{{\rm\,km\,s^{-1}}}
\def\kpc{{\rm\,kpc}}
\def\Mpc{{\rm\,Mpc}}

\def\msun{{\rm\,M_\odot}}
\def\lsun{{\rm\,L_\odot}}

\def\pc{{\rm\,pc}}

\def\Gyr{{\rm\,Gyr}}

\def\annrev #1 #2 {ARA\&A, #1, #2}
\def\aa #1 #2 {A\&A, #1, #2}
\def\aasupp #1 #2 {A\&AS, #1, #2}
\def\aj #1 #2 {AJ, #1, #2}
\def\apj #1 #2 {ApJ, #1, #2}
\def\apjlett #1 #2 {ApJ, #1, #2}
\def\apjsupp #1 #2 {ApJS, #1, #2}
\def\ban #1 #2 {Bull.\ Astron.\ Inst.\ Netherlands, #1, #2}
\def\mn #1 #2 {MNRAS, #1, #2}
\def\nature #1 #2 {Nat, #1, #2}
\def\pasj #1 #2 {PASJ, #1, #2}
\def\pasp #1 #2 {PASP, #1, #2}